\documentclass[3p,times]{elsarticle}

\usepackage{ecrc}

\volume{00}

\firstpage{1}

\journalname{Nuclear Physics A}

\runauth{J. M. Durham (for the PHENIX collaboration)}

\jid{npa}

\jnltitlelogo{Nuclear Physics A}

\usepackage{amssymb}

\biboptions{square,comma,numbers,sort&compress}

\usepackage[figuresright]{rotating}

\begin{document}

\begin{frontmatter}



\dochead{}

\title{PHENIX Results on Heavy Quarks at Low $x$}


\author{J. Matthew Durham\corref{cor}}
\author{\\for the PHENIX Collaboration}
\address{Los Alamos National Laboratory, Los Alamos, NM 87545, USA}

\cortext[cor]{durham@lanl.gov}

\begin{abstract}
It is becoming increasingly clear that initial state effects inherent to collisions of nuclei play an important role in the interpretation of data from heavy ion collisions at RHIC and the LHC. Such effects are more apparent in kinematic regions where the gluon density is expected to be significantly modified in the nucleus. The PHENIX experiment has studied these effects through the production of heavy quarks at backwards, middle, and forward rapidity, where partonic interactions in the nucleus and changes in the gluon structure function influence heavy quark production in different ways. Comparisons between these different rapidities in $d+$Au collisions offer us a window into the dynamics of particle production and transport in the nucleus. In these proceedings, new PHENIX results on heavy quark production at low $x$ values are discussed, in the context of A+A data from RHIC and the LHC.
\end{abstract}





\end{frontmatter}


\section{Introduction}
\label{intro}

Measurements of heavy quark production in nuclear collisions have been a focus of the PHENIX experiment since the beginning of the RHIC program.  The relatively large mass of heavy quarks ensures that they are only produced in the early stages of the collision, rather than through thermal excitation as the medium evolves, and the production mechanism is dominated by processes which are calculable with perturbative QCD techniques \cite{pp_fit, PPG077}.  As such, open heavy flavor provides a relatively well calibrated probe of medium effects that are present in nuclear collisions.  PHENIX has the ability to measure heavy quarks through their decays to electrons at midrapidity and to muons at forward rapidity \cite{PHENIXNIM}.  

Several factors will affect charm and bottom production in nuclear collisions:  Since the dominant production mechanism for heavy quarks at RHIC is gluon fusion, modifications of the gluon structure function in nuclei will directly affect charm and bottom production rates.  While passing through the nucleus, interactions between the target and nuclear constituents may lead to broadening of the $p_{T}$ spectrum of the observed final state hadrons, as well as induce energy loss of colored partons.  The Cronin effect that has been observed for light flavor hadrons may also influence charmed and bottom hadrons \cite{PPG146}.  However, the mechanism behind the Cronin effect is not immediately clear.  If the $p_{T}$ broadening is due to parton scattering in the nucleus, it may be mass dependent, in which case the heavy charm and bottom quarks will be especially sensitive.  If the mechanism is quark recombination, influenced by effects that may be of hydrodynamic origin \cite{CMS_pPb_flow, ALICE_pPb_flow, ATLAS_pPb_flow, PPG152}, then heavy flavor baryons may be enhanced over mesons.

\section{Open Heavy Flavor in $d$+Au collisions}

Colliding deuterons with large nuclei allows heavy quark production in a nuclear target to be studied with minimal effects from the deconfined medium that is prevalent in A+A collisions.  The PHENIX collaboration has previously shown that electrons from decays of open heavy flavor hadrons are enhanced at midrapidity in central $d+$Au collisions, in contrast to the large suppression that is observed in Au+Au collisions at the same energy \cite{PPG131}.  Since these decay products are measured rather than fully reconstructed hadronic states, it is not possible to separate out hadrons containing charm and bottom, or to distinguish between contributions from mesons and baryons.       

\begin{figure}
\centering
\begin{minipage}{.3\textwidth}
\centering
\includegraphics[trim = 0mm 0mm 0mm 3mm, clip, width=\linewidth]{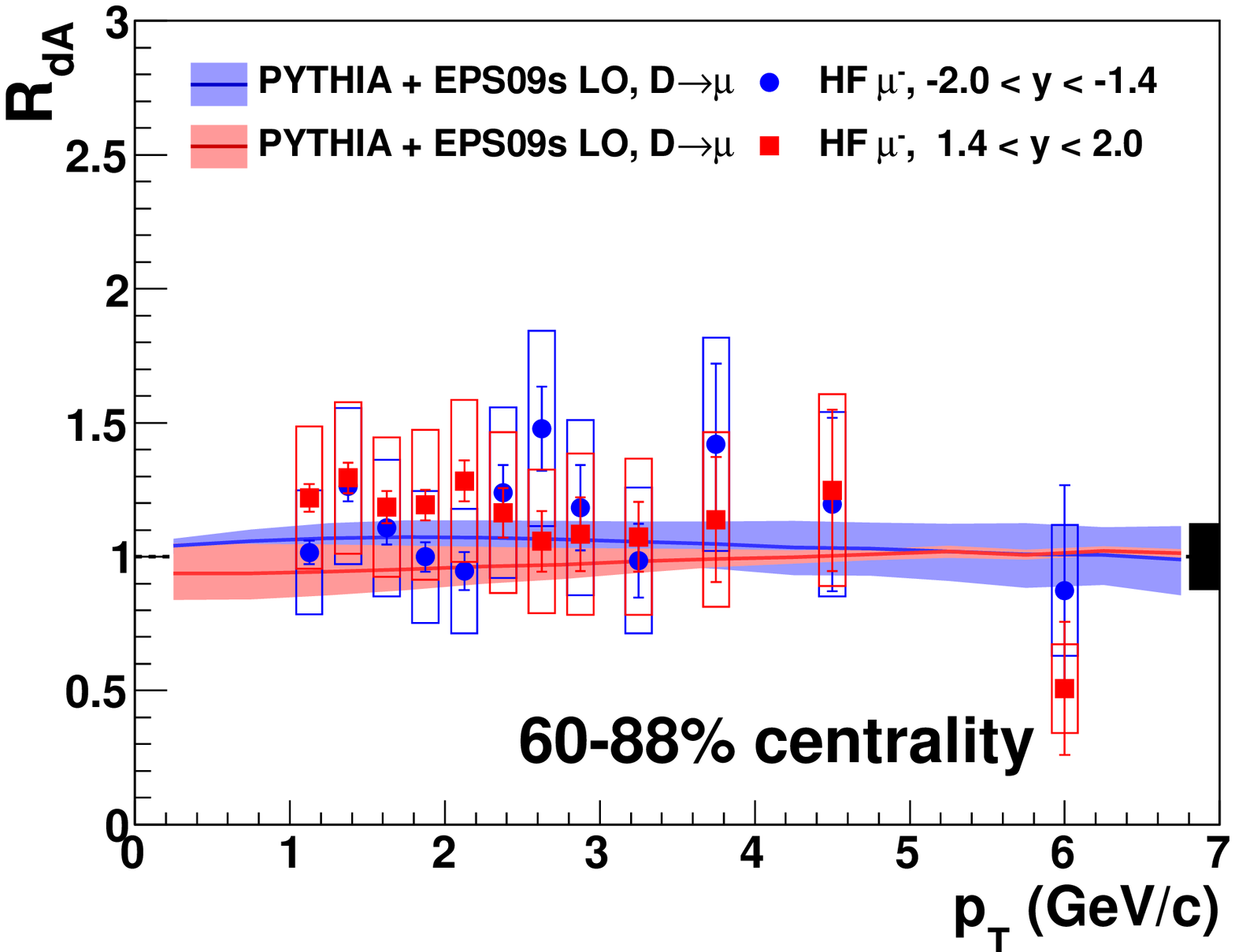}
\caption{$R_{dA}$ for heavy flavor muons in peripheral collisions, which is consistent with one, and with a calculation based on the EPS09s PDF set.}
\label{fig:RdA_periph}
\end{minipage}\hfill
\begin{minipage}{.3\textwidth}
\centering
\includegraphics[trim = 0mm 0mm 0mm 5mm, clip, width=\linewidth]{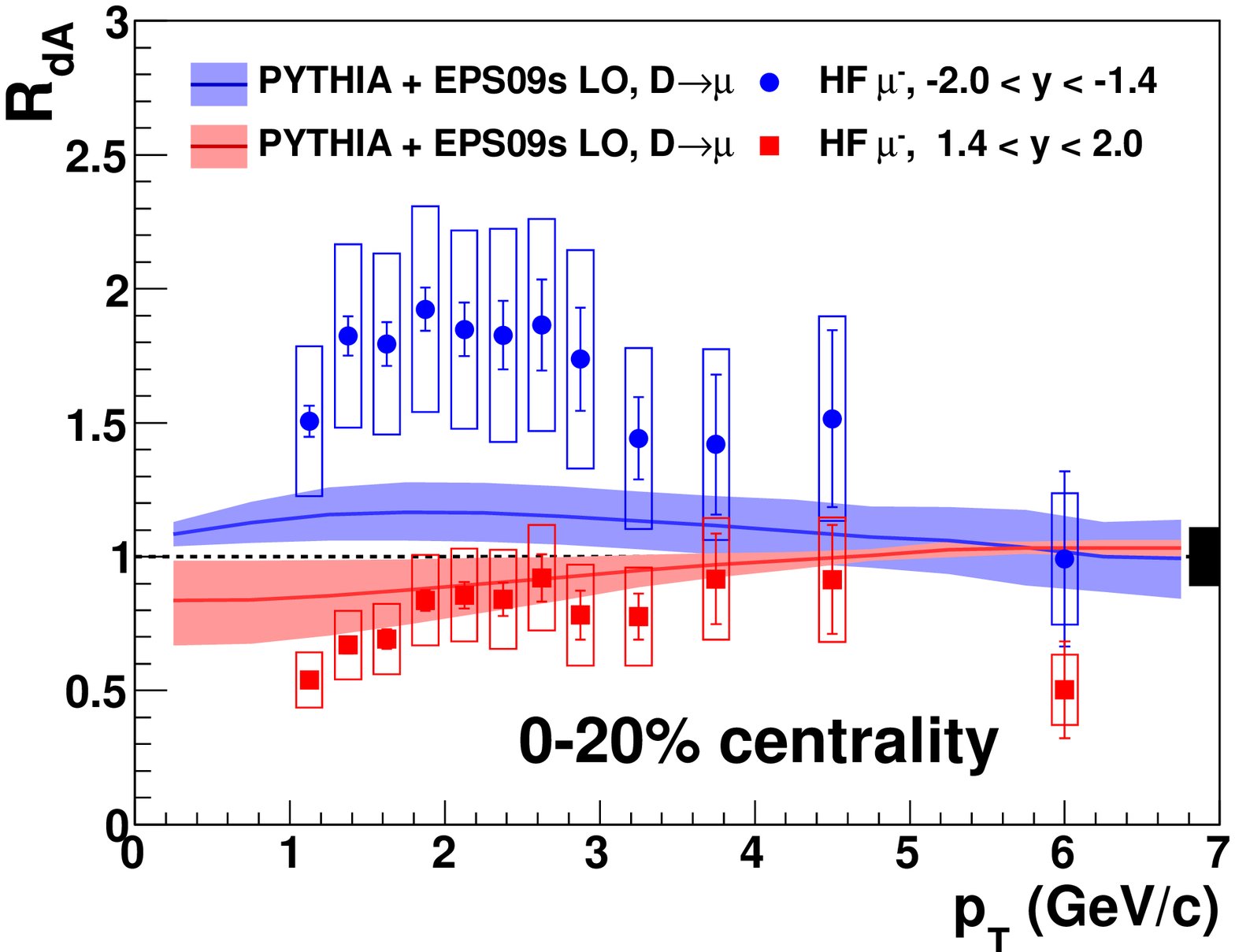}
\caption{$R_{dA}$ for heavy flavor muons in central collisions.  The backward rapidity data shows enhancement beyond what is expected from anti-shadowing.}
\label{fig:RdA_cent}
\end{minipage}\hfill
\begin{minipage}{.3\textwidth}
\centering
\includegraphics[width=\linewidth]{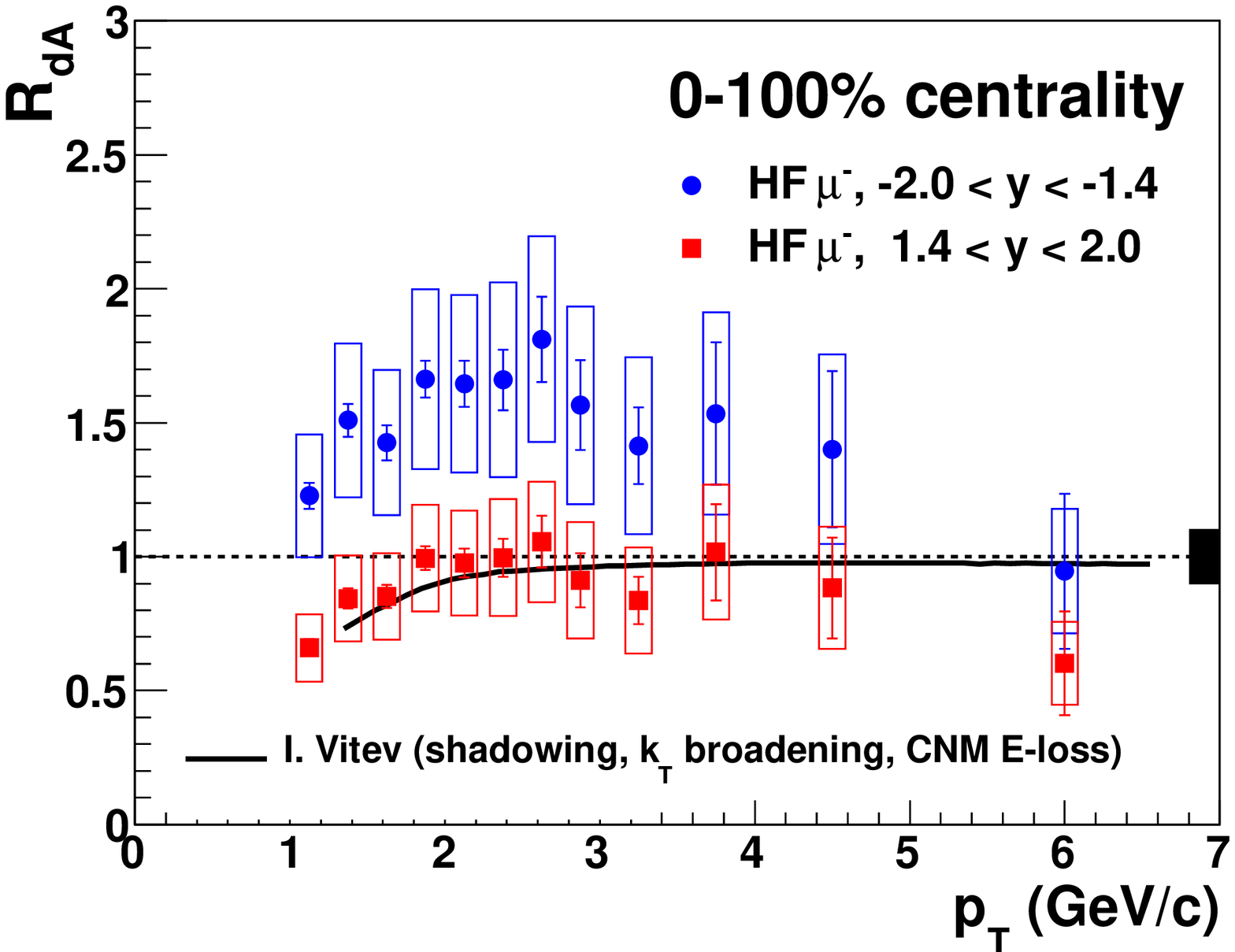}
\caption{$R_{dA}$ for heavy flavor muons in Minimum Bias collisions.  The forward rapidity data is consistent with a realistic calculation from \cite{Vitev_HF}.}
\label{fig:RdA_MB}
\end{minipage}
\end{figure}

Charm production at midrapidity at PHENIX samples $x$ values in the low to mid $10^{-2}$ range, near the crossover between gluon shadowing and anti-shadowing.  This measurement has been extended to forward and backward rapidity \cite{PPG153}, allowing $x$ values in the mid  $10^{-3}$ (shadowing) and high $10^{-2}$ (anti-shadowing) range to be examined.  Fig. \ref{fig:RdA_periph} shows the nuclear modification factors $R_{dA}$ for muons from open heavy flavor decays in peripheral $d+$Au collisions at forward and backward rapidity, which are consistent with a calculation based on the EPS09s PDF set \cite{EPS09s} for muons from $D$ meson decay, and with one.  The central $R_{dA}$ shown in Fig. \ref{fig:RdA_cent} is consistent with the EPS09s calculation at forward rapidity, however, at backward rapidity the observed enhancement is larger than is expected from anti-shadowing alone.  This may be due to interactions with co-moving particles, which are more prevalent in the Au-going direction, or due to enhancement of the baryon sample (which is present in the data but not included in the calculation).  The forward Minimum Bias data shown in Fig. \ref{fig:RdA_MB} is consistent with a realistic calculation that includes shadowing, a Cronin effect due to parton scattering, and energy loss in the nucleus \cite{Vitev_HF}.  These measurements imply that shadowing contributes to heavy quark suppression at forward rapidity in heavy ion collisions.

\section{Open Heavy Flavor in $d+$Au as a Quarkonia Baseline}
\label{hidden}
A great amount of effort been devoted to explaining quarkonia suppression in nuclear collisions ( cf. \cite{Jpsi_theory_review, Vitev_Jpsi, McGlinchey_Jpsi} and references therein).  Measurements at fixed target experiments and in $d$+Au collisions at RHIC \cite{E866, NA50_Jpsi, NA60_Jpsi, PPG125} have shown that $J/\psi$ production is suppressed in collisions where effects from a deconfined medium are expected to be minimal, which suggests that a mechanism other than color screening is inhibiting bound state $c \bar{c}$ production.  While shadowing, energy loss, and the Cronin effect can modify production of all hadrons containing charm, only the bound states are subject to breakup effects, which may occur while the $c\bar{c}$ precursor state is in the nucleus or through interactions with co-moving particles.  As previously noted, none of these nuclear effects are present in $p+p$ collisions, so interpretations of charmonia suppression relying on a production baseline from that system require theoretical models to disentangle the various effects.  However, a comparison of the relative suppression of hidden and open heavy flavor in nuclear collisions is sensitive only to the magnitude of charmonia breakup.  The naive expectation is that $R_{dA}$ of charmonia can only be less than $R_{dA}$ of open charm, since only charmonia is subject to this additional suppression mechanism.  

\begin{figure}
\centering
\begin{minipage}{.29\textwidth}
\centering
\includegraphics[width=\linewidth]{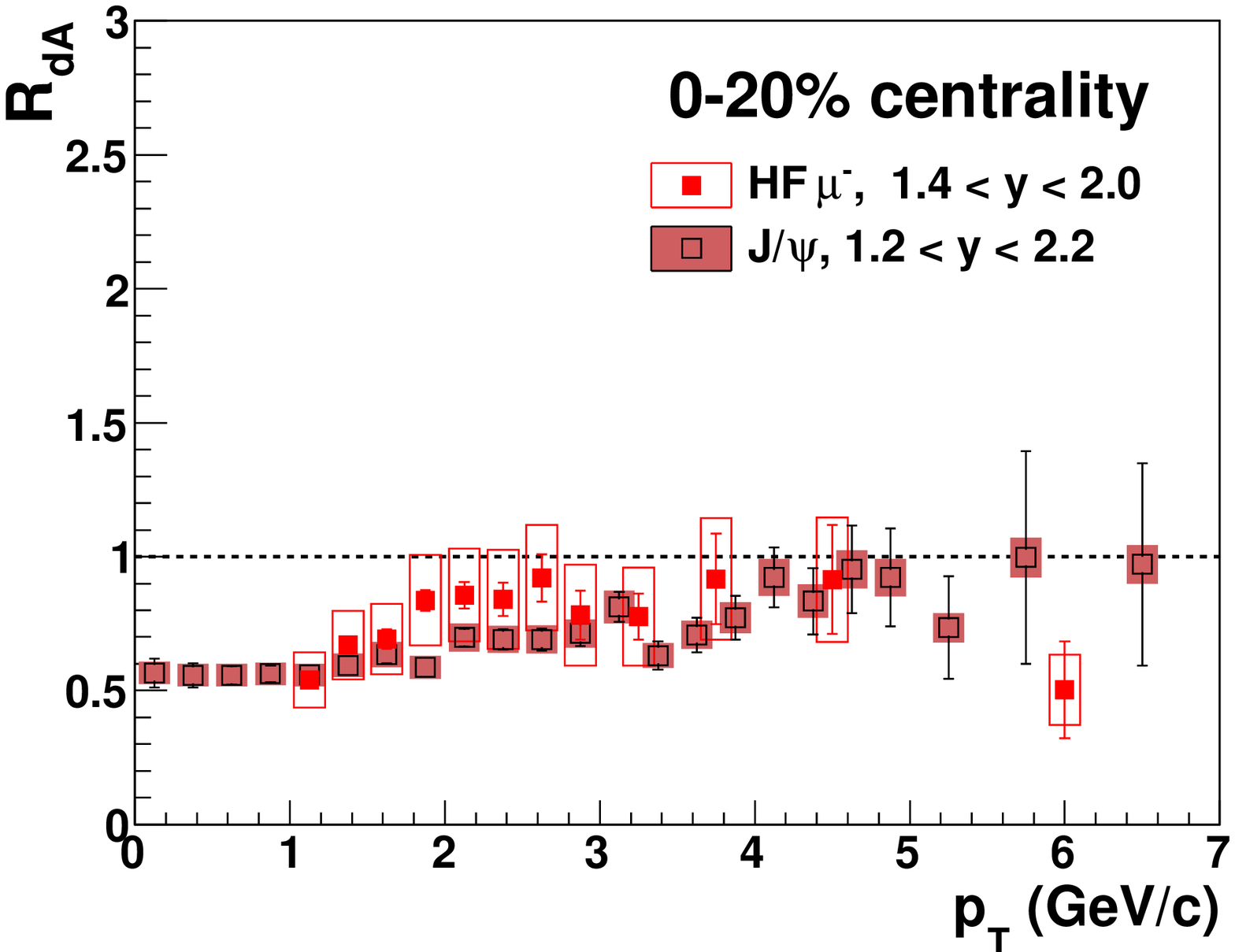}
\caption{$R_{dA}$ of open and hidden heavy flavor at backward rapidity ($d-$going direction).}
\label{fig:fwd}
\end{minipage}\hfill
\begin{minipage}{.33\textwidth}
\centering
\includegraphics[trim = 0mm 0mm 0mm 12mm, clip, width=\linewidth]{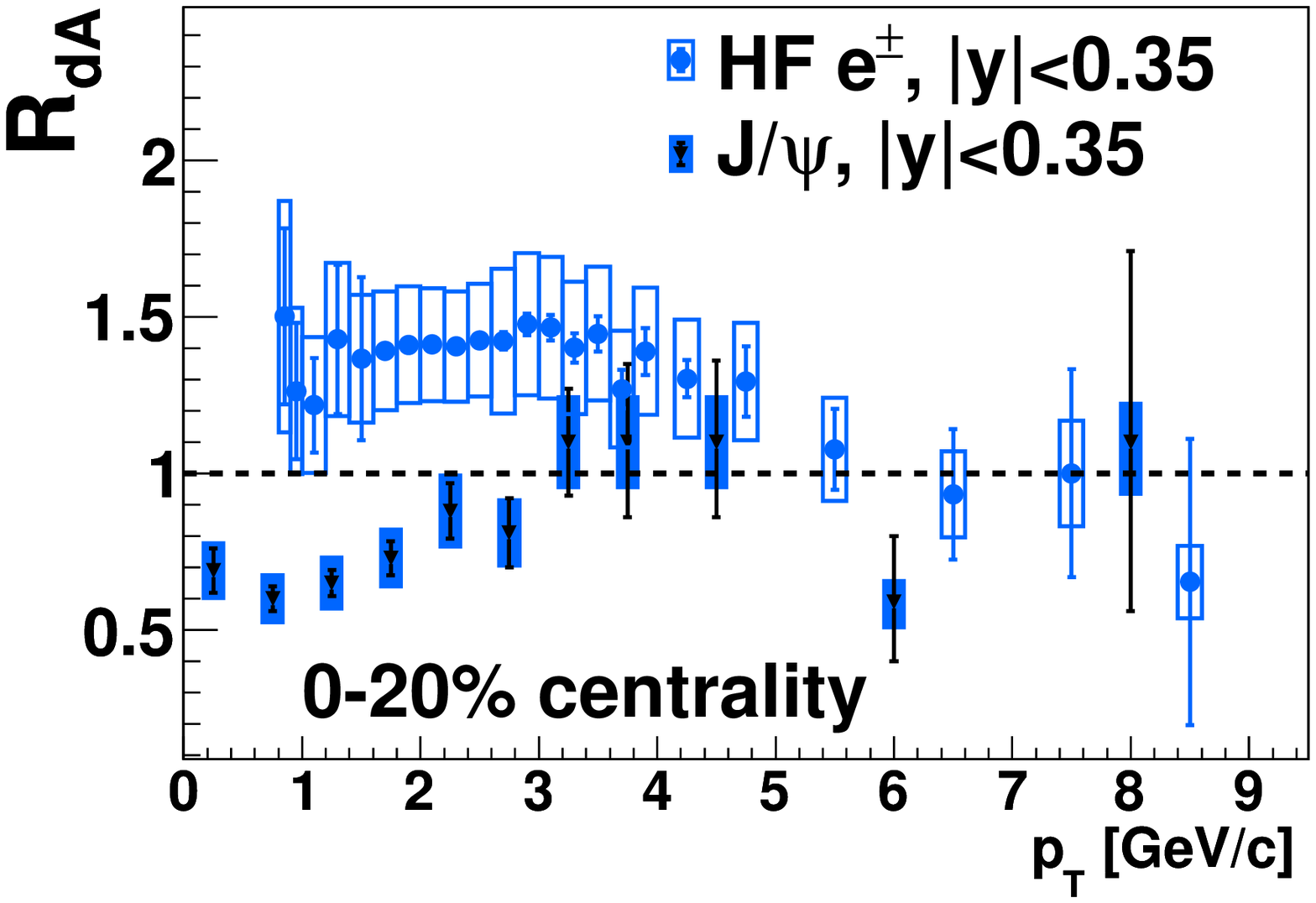}
\caption{$R_{dA}$ of open and hidden heavy flavor at midrapidity.}
\label{fig:mid}
\end{minipage}\hfill
\begin{minipage}{.29\textwidth}
\centering
\includegraphics[width=\linewidth]{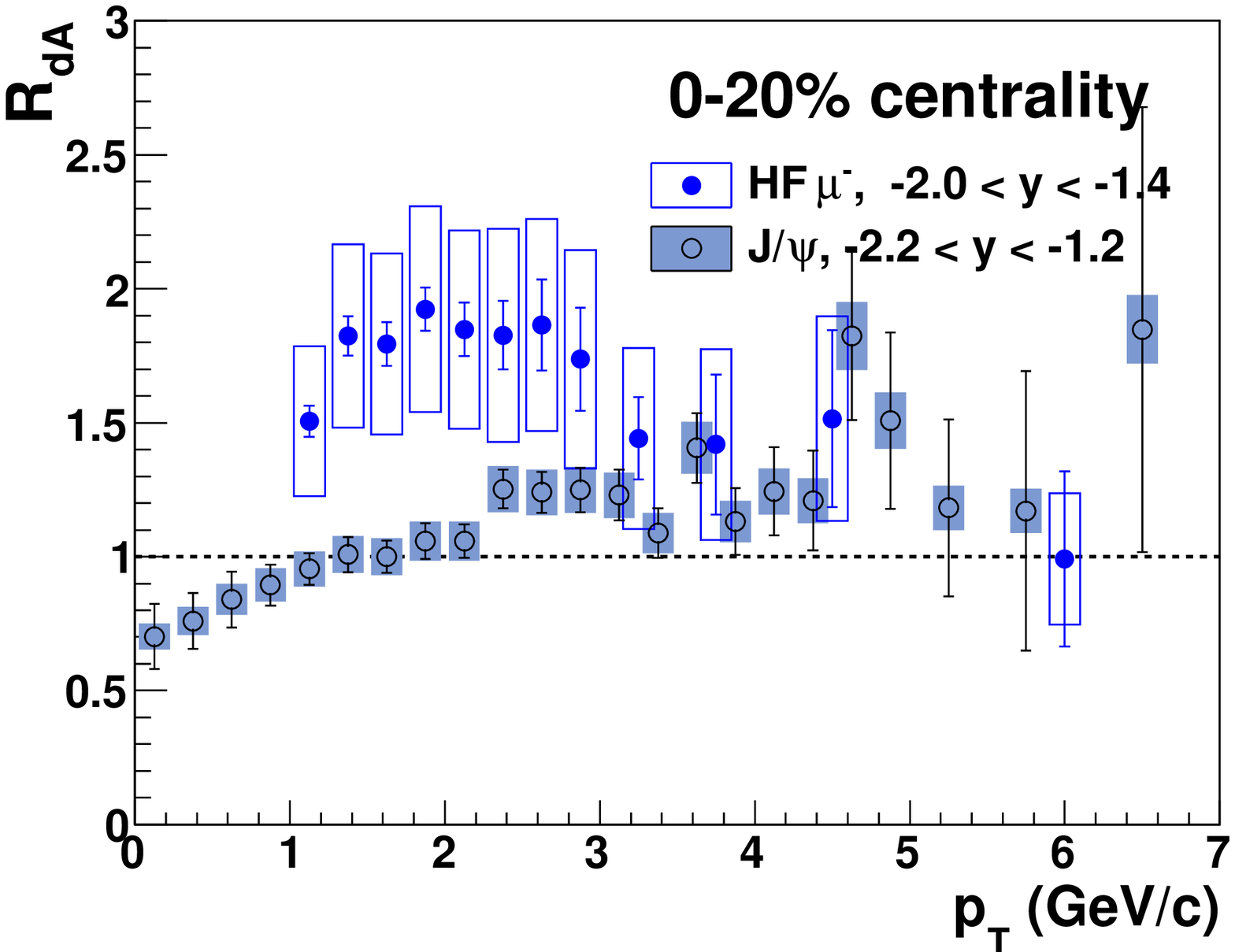}
\caption{$R_{dA}$ of open and hidden heavy flavor at forward rapidity (Au-going direction).}
\label{fig:bkwd}
\end{minipage}
\end{figure}

Figs. \ref{fig:fwd} through \ref{fig:bkwd} show the $p_{T}$ dependence of the nuclear modification factor $R_{dA}$ for fully reconstructed $J/\psi$ mesons and leptons from open heavy flavor decays, separated into three rapidity ranges.  At forward rapidity (left panel) a similar behavior is observed for both open heavy flavor  and $J/\psi$ mesons.  In this region, the density of produced co-moving particles is small and the time spent in the nucleus is relatively short.  At mid- and backward rapidity, however, a markedly different behavior is shown for open and hidden heavy flavor: open heavy flavor is enhanced while the $J/\psi$ is suppressed.  In these regions, the rapidity density of produced particles is higher than at forward rapidity, so the probability of an interaction between the $J/\psi$ and a co-moving particle (which may be a break-up mechanism) is more likely.  In addition, the time the $c\bar{c}$ pair spends in the nucleus is longer, so dissociation through interactions with the nuclear remnant can occur more frequently.  However, we note here that the kinematic differences between fully reconstructed $J/\psi$ mesons and leptons from decays of hadrons containing a single charm quark require careful consideration before conclusions can be drawn about the underlying charm quark dynamics.

\section{Open Heavy Flavor in Cu+Cu collisions}

 The flexibility of RHIC has allowed the PHENIX collaboration to make high-statistics measurements of heavy flavor production in Cu+Cu collisions at $\sqrt{s_{NN}}$ = 200 GeV \cite{PPG150}.  This intermediately-sized system permits the transition region between the small $d+$Au system (where open heavy flavor is enhanced) and Au+Au collisions (where heavy quarks are dramatically suppressed) to be studied in detail.

\begin{figure}[htbp]
\centering
\begin{minipage}{.5\textwidth}
\centering
\includegraphics[width=0.75\linewidth]{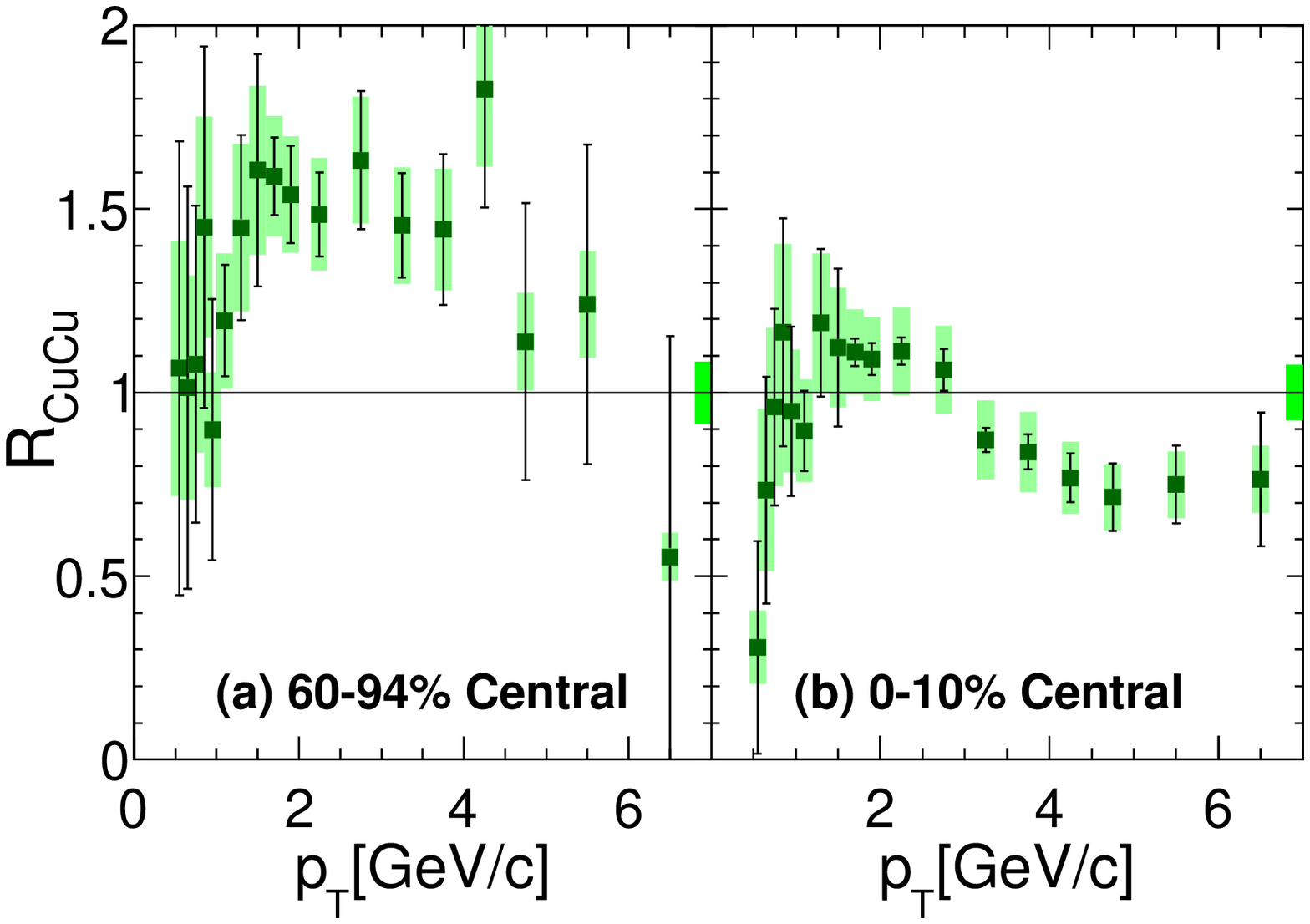}
\caption{$R_{AA}$ for electrons from open heavy flavor decays at midrapidity, in a) peripheral and b) central Cu+Cu collisions.}
\label{fig:Cu}
\end{minipage}\hfill
\begin{minipage}{.5\textwidth}
\centering
\includegraphics[trim = 0mm 0mm 0mm 9mm, clip, width=0.75\linewidth]{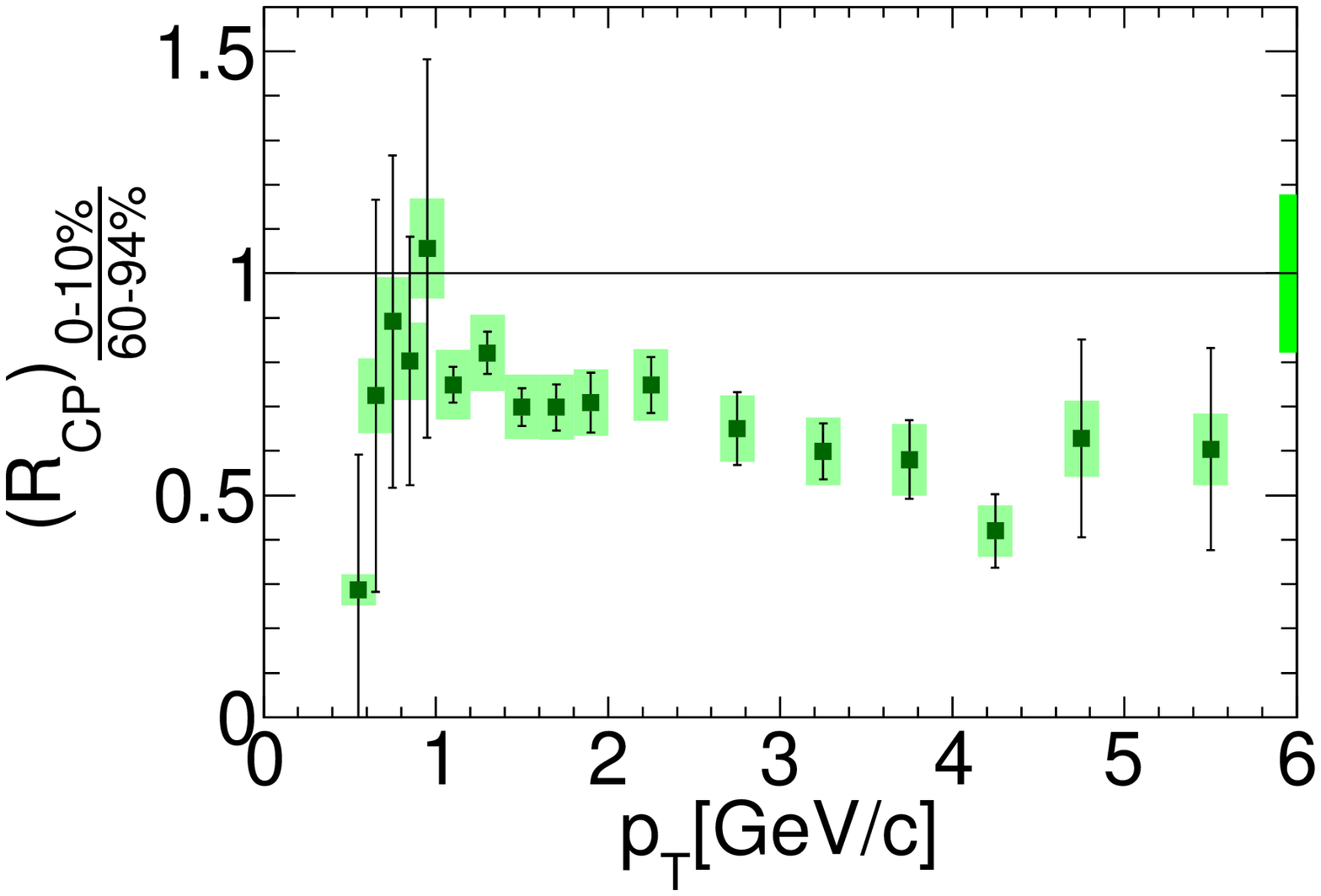}
\caption{$R_{cp}$ for open heavy flavor electrons in Cu+Cu collisions.}
\label{fig:Cu_Rcp}
\end{minipage}
\end{figure}

Fig. \ref{fig:Cu} shows the nuclear modification factor for electrons from open heavy flavor decays in peripheral and central Cu+Cu collisions.  In peripheral collisions, a surprising enhancement of open heavy flavor is observed, while in central collisions a moderate suppression is found for $p_{T} >$ 3 GeV/$c$.  As the $p+p$ reference data used to calculate $R_{AA}$ does not account for the enhancement observed in peripheral collisions, the ratio $R_{cp}$ is shown in Fig. \ref{fig:Cu_Rcp}.  Here we see that significant suppression emerges within the Cu+Cu collision system, presumably due to energy loss effects in central collisions.

\begin{figure}[htbp]
  \centering
  \includegraphics[width=0.8\linewidth]{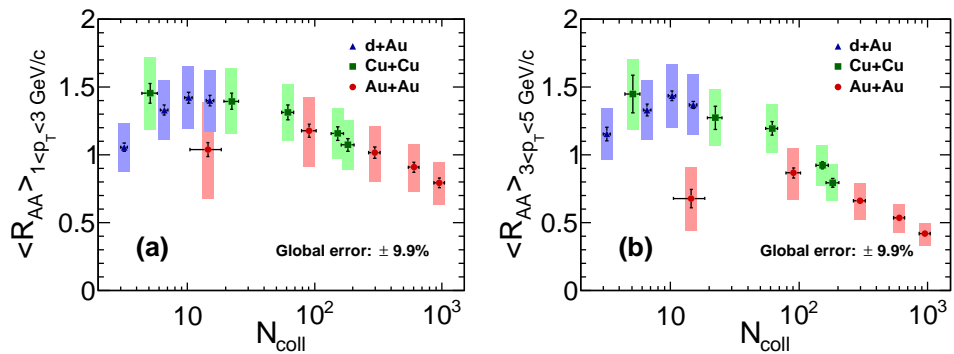}
  \caption{The midrapidity heavy flavor electron $R_{AA}$ as a function of $N_{coll}$ for $d+$Au, Cu+Cu, and Au+Au collisions at RHIC.}
  \label{fig:RAA_Ncoll}
\end{figure}

A comparison between the modification of open heavy flavor in $d+$Au, Cu+Cu, and Au+Au collisions as a function of $N_{coll}$ is shown in Fig. \ref{fig:RAA_Ncoll}, for two momentum ranges.  The $d+$Au data shows an increasing enhancement with $N_{coll}$, while the Au+Au data shows increasing suppression.  The Cu+Cu data is consistent with these systems at similar $N_{coll}$ values, and displays a transition from enhancement to suppression with increasing centrality.

\section{Summary and Future Prospects}
\label{future}

Heavy quark production in a nuclear target is sensitive to a variety of effects which, depending on the kinematic range sampled, can enhance or suppress production relative to a $p+p$ baseline.  In $d+$Au collisions at forward rapidity, where gluon shadowing is expected to be significant, all heavy flavor production is suppressed at low $p_{T}$.  At mid- and forward rapidity, open heavy flavor is enhanced while $J/\psi$ production is suppressed, providing direct evidence of a significant charmonia breakup effect.  As the system size increases, the open heavy flavor enhancement mechanism is gradually overtaken by suppression in central Cu+Cu and semi-peripheral Au+Au collisions, presumably due to the emerging dominance of energy loss in the deconfined nuclear medium.

The PHENIX collaboration has recently installed a new silicon tracking detector, the FVTX,  which provides precise charged particle tracking at forward and backward rapidity \cite{FVTXNIM}.  The new capabilities provided by this detector will allow measurements of open charm and open bottom to be made separately in nuclear collisions.  The PHENIX collaboration has requested a scan of nuclear targets in $p+$A collisions in 2015, where A = C, Cu, and Au; this will allow effects sensitive to collision geometry and the impact parameter dependence of gluon PDF modifications to be examined in detail.





\bibliographystyle{elsarticle-num}
\bibliography{ecrc-templateHP}







\end{document}